\documentstyle[11pt,amssymb,epsf]{article}

\def\ben{\begin{equation}}
\def\een{\end{equation}}

\let\a=\alpha    
    
  \let\n=\nu

\let\C=\Chi

\def\nn{\nonumber} \def\bd{\begin{document}} \def\ed{\end{document}}
\def\ds{\documentstyle} \let\fr=\frac \let\bl=\bigl \let\br=\bigr
\let\Br=\Bigr \let\Bl=\Bigl
\let\bm=\bibitem
\let\na=\nabla
\let\pa=\partial \let\ov=\overline
\newcommand{\be}{\begin{equation}}
\newcommand{\ee}{\end{equation}}
\def\ba{\begin{array}}
\def\ea{\end{array}}
\def\ft#1#2{{\textstyle{{\scriptstyle #1}\over {\scriptstyle #2}}}}
\def\fft#1#2{{#1 \over #2}}
\def\del{\partial}
\def\vp{\varphi}
\def\sst#1{{\scriptscriptstyle #1}}
\def\oneone{\rlap 1\mkern4mu{\rm l}}
\def\td{\tilde}
\def\wtd{\widetilde}
\def\ie{\rm i.e.\ }
\def\dalemb#1#2{{\vbox{\hrule height .#2pt
        \hbox{\vrule width.#2pt height#1pt \kern#1pt
                \vrule width.#2pt}
        \hrule height.#2pt}}}
\def\square{\mathord{\dalemb{6.8}{7}\hbox{\hskip1pt}}}
\newcommand{\ho}[1]{$\, ^{#1}$}
\newcommand{\hoch}[1]{$\, ^{#1}$}
\newcommand{\bea}{\begin{eqnarray}}
\newcommand{\eea}{\end{eqnarray}}
\newcommand{\ra}{\rightarrow}
\newcommand{\lra}{\longrightarrow}
\newcommand{\Lra}{\Leftrightarrow}
\newcommand{\ap}{\alpha^\prime}
\newcommand{\bp}{\tilde \beta^\prime}
\newcommand{\tr}{{\rm tr} }
\newcommand{\Tr}{{\rm Tr} }
\def\0{{\sst{(0)}}}
\def\1{{\sst{(1)}}}
\def\2{{\sst{(2)}}}
\def\3{{\sst{(3)}}}
\def\4{{\sst{(4)}}}
\def\5{{\sst{(5)}}}
\def\6{{\sst{(6)}}}
\def\7{{\sst{(7)}}}
\def\8{{\sst{(8)}}}
\def\n{{\sst{(n)}}}
\def\cA{{{\cal A}}}
\def\cB{{{\cal B}}}
\def\cF{{{\cal F}}}
\def\tV{\widetilde V}
\def\tW{\widetilde W}
\def\tH{\widetilde H}
\def\tE{\widetilde E}
\def\tF{\widetilde F}
\def\tA{\widetilde A}
\def\im{{{\rm i}}}
\def\tY{{{\wtd Y}}}
\def\ep{{\epsilon}}
\def\vep{{\varepsilon}}
\def\R{\rlap{\rm I}\mkern3mu{\rm R}}
\def\bD{{{\bar D}}}

\def\R{\rlap{\rm I}\mkern3mu{\rm R}}
\def\bD{{{\bar D}}}
\def\R{{{\Bbb R}}}
\def\C{{{\Bbb C}}}
\def\H{{{\Bbb H}}}
\def\CP{{{\Bbb C}{\Bbb P}}}
\def\RP{{{\Bbb R}{\Bbb P}}}
\def\Z{{{\Bbb Z}}}
\def\bA{{{\Bbb A}}}
\def\bB{{{\Bbb B}}}
\def\bC{{{\Bbb C}}}
\def\bD{{{\Bbb D}}}
\def\bE{{{\Bbb E}}}
\def\bZ{{{\Bbb Z}}}
\def\Re{{{\frak{Re}}}}
\def\Im{{{\frak{Im}}}}
\def\cosec{{\,\hbox{cosec}\,}}
\def\Gm{{\Gamma_{\!\! -}}}
\def\Gp{{\Gamma_{\!\! +}}}
\def\stan{{standard }}
\def\nonstan{{supernumerary }}

\newcommand{\tamphys}{\it Center for Theoretical Physics,
Texas A\&M University, College Station, TX 77843}

\newcommand{\upenn}{\it Department of Physics and Astronomy,\\ University
of Pennsylvania, Philadelphia, PA 19104}

\newcommand{\brussels}{\it Physique Th\'eorique et Math\'ematique, 
Universit\'e Libre de Bruxelles,\\ Campus Plaine C.P. 231, B-1050
Bruxelles, Belgium} 

\newcommand{\auth}{H. L\"u\hoch{\dagger1} and J.F. 
V\'azquez-Poritz\hoch{\ddagger2}}

\begin{document}
\begin{flushright}
MCTP-02-19\\
ULB-TH/02-06\\
March  2002\\
\hfill{\bf hep-th/0204001}\\
\end{flushright}

\vspace{10pt}

\begin{center}

{\large {\bf Penrose Limits of Non-standard Brane Intersections}}

\vspace{20pt}
\auth

\vspace{10pt}
{\hoch{\dagger}\it Michigan Center for Theoretical Physics\\
University of Michigan, Ann Arbor, Michigan 48109}

\vspace{10pt}
{\hoch{\ddagger}\brussels}

\vspace{30pt}

\underline{ABSTRACT}
\end{center}

       The non-standard intersection of two 5-branes and a string can
give rise to AdS$_3\times S^3\times S^3\times S^1$.  We consider the
Penrose limit of this geometry and study the supersymmetry of the
resulting pp-wave solution.  There is a one-parameter family of
Penrose limits associated with the orthogonal rotation of the two
foliating circles within the two 3-spheres.  Supernumerary Killing
spinors arise only when the rotation angle is 45 degrees, for which
case we obtain the corresponding light-cone string action that has
linearly-realised supersymmetry.  We also obtain Penrose limits of
other non-standard intersections that give rise to the product of
AdS$_3$ or AdS$_2$ and two spheres.  The resulting pp-waves are
supported by multiple constant field strengths.

{\vfill\leftline{}\vfill
\vskip 10pt 
\footnoterule {\footnotesize \hoch{1} 
Research supported in full by DOE grant DE-FG02-95ER40899.

{\footnotesize \hoch{2} Research supported in full by the Francqui
Foundation (Belgium), the Actions de Recherche Concert{\'e}es 
\phantom{of the} of the
Direction de la Recherche Scientifique - Communaut\'e Francaise de
Belgique, IISN-Belgium
\phantom{of the} (convention 4.4505.86).  }
\vskip  -12pt}

\pagebreak
\setcounter{page}{1}

\section{Introduction}

    PP-waves arising from the Penrose limits of AdS$_p\times S^q$
spacetime provide simple backgrounds on which to study string and
M-theory.  The maximally supersymmetric pp-wave with a constant 4-form
field strength was first constructed in \cite{kg} for
eleven-dimensional supergravity.  Recently, the maximally
supersymmetric pp-wave in type IIB supergravity with a constant
self-dual 5-form field strength was obtained in \cite{blafighulpap}.
These solutions are actually the Penrose limits of $AdS_p\times S^q$
spacetimes with $(p,q)=(4,7), (7,4)$ and (5,5) respectively
\cite{blafighulpap2}.  In the type IIB case, the light-cone string
action for the pp-wave \cite{met,bermalnas}, is exactly solvable.
This makes it possible to study the AdS/CFT correspondence on the
level of full string theory in the Penrose limit \cite{bermalnas},
which has generated considerable interest recently
\cite{bur1}-\cite{bur13}.

      AdS$_p \times S^q$ arises as the near-horizon geometry of
non-dilatonic branes or intersecting branes, for $(p,q)=(4,7)$,
$(7,4)$, $(5,5)$, $(3,3)$, $(3,2)$, $(2,3)$ and $(2,2)$.  AdS
structure also arises from ``non-standard'' intersections, for which
the harmonic function $H$ for each brane component depends on the
coordinates of the relative transverse space rather than those of the
overall transverse space \cite{khuri,bbj,gkt}.  The simplest example
is the non-standard intersection of two 5-branes plus a string in the
common worldvolume, whose near-horizon geometry is AdS$_3\times
S^3\times S^3\times S^1$ \cite{cow}.

      In section 2, we perform the Penrose limit on the above
solution.  Since the product space involves two 3-spheres, there
exists a one-parameter family of Penrose limits of this solution,
parametrized by the rotation of the two foliating $S^1$ within the two
3-spheres.  In section 3, we study the supersymmetry of the resulting
pp-wave solution.  We find that there are 16 standard Killing spinors
associated with any pp-wave solution, but supernumerary Killing
spinors arise only for a specific choice of the parameter, associated
with a 45 degree rotation of the two $S^1$.  For this case, there are
4 supernumerary Killing spinors, which are all independent of the
$x^+$ coordinate.  Furthermore, 4 out of the 16 standard Killing
spinors are also independent of $x^+$.  We then obtain the
supersymmetric light-cone string action in this pp-wave background.
In section 4, we T-dualize the pp-wave solution and show that it is
related to the D3-brane with one of the world-volume coordinates
fibred over the six-dimensional transverse space.  In section 5, we
obtain the Penrose limits of more examples of non-standard
intersections which give rise to AdS$_2$ or AdS$_3$.  Again, all of
these examples have a one-parameter family of Penrose limits.  We
conclude our paper in section 6.

\section{5-5-1 system and Penrose limit}

    The 5-5-1 system is supported by a 3-form field strength in
$D=10$.  The relevant Lagrangian is given by
\be 
e^{-1}\, {\cal L} = R - \ft12(\del\phi)^2 -\ft{1}{12}
{\rm e}^{-\phi}\, F_\3^2\,.
\ee
Since our system involves only the metric, the dilaton and the 3-form
field strength, the results apply to types IIA, IIB and heterotic
string theories.  The solution is given by \cite{cow}
\bea
ds_{10}^2&=&K^{-3/4}(H\,\wtd H)^{-1/4}(-dt^2+dx^2+K\,H\,dy_i^2+ K\,{\wtd
H}\,d\td y_i^2)\,,\nn\\
F_\3 &=& {\rm e}^{\phi} {\ast({\wtd H}\ dt\wedge dx\wedge d^4 \td y
\wedge dH^{-1})}+ {\rm e}^{\phi}{\ast(H\ dt\wedge dx\wedge d^4 y
\wedge d{\wtd H}^{-1})}\\
&&+dt \wedge dx\wedge dK^{-1}\,,\quad
\phi= -\frac{1}{2}{\rm log}[K/(H{\wtd H})]\,,\qquad K=H\,\wtd H\nn
\eea
where $H$ and $\wtd H$ are harmonic functions in the relative
transverse $y_i$ and $\td y_i$ spaces respectively.
The metric can be represented by the following diagram:

\bigskip\bigskip
\centerline{
\begin{tabular}{c|ccccccccccc}
&$t$ & $x$ & $y_1$ & $y_2$ & $y_3$ & $y_4$ & $\td y_1$ & 
$\td y_2$ & $\td y_3$ & $\td y_4$ & \\ \hline
5-brane&$\times$ & $\times$ & $\times$ & $\times$ & $\times$ & 
$\times$ & $-$ & $-$ & $-$ & $-$ & ${\tilde H}$ \\
5-brane&$\times$ & $\times$ & $-$ & $-$ & $-$ & $-$ & 
$\times$ & $\times$ & $\times$ & $\times$ & $H$ \\
string&$\times$ & $\times$ & $-$ & $-$ & $-$ & $-$ & 
$-$ & $-$ & $-$ & $-$ & $K$ \\
\end{tabular}}
\bigskip

\centerline{Diagram 1. 5-5-1 system}
\bigskip\bigskip

We parametrize the coordinates as $dy_i^2=dy^2+y^2\,
d\Omega_3^2$ and $d\td y_i^2=dy^2+y^2\, d{\wtd \Omega}_3^2$.  The
isotropic solution is then given by
\be
H=1+\frac{Q}{y^2}\,,\qquad
{\wtd H}=1+\frac{{\wtd Q}}{\td y^2}\,.\label{H4}
\ee
For simplicity, we take $Q={\wtd Q}=\lambda^2$.  The near-horizon
geometry of the solution is $AdS_3\times S^3\times S^3\times S^1$
\cite{cow}.  Expressing the AdS$_3$ in global coordinates, and each
$S^3$ as a foliation of two circles, we have
\bea
ds_{10}^2&=& d\varphi^2 + \lambda^2\,\Big(\ft12(-\cosh^2\rho\, dt^2 +
d\rho^2+{\sinh}^2\rho\,
d\gamma^2)\label{ads3s3s3met}\\
&&+(\cos^2 \theta\, d\psi^2+d\theta^2+\sin^2\theta\, d\phi^2)
+(\cos^2 {\td \theta}\, d{\td \psi}^2+d{\td \theta}^2+
\sin^2{\td \theta}\,d{\td \phi}^2)\Big)\,,\nn\\
F_\3&=&\lambda^2\, (\epsilon_\3 +2\Omega_\3 + 2\wtd \Omega_\3)\,,
\label{ad3s3s3f3}
\eea
where $\epsilon_\3$, $\Omega_\3$ and $\wtd \Omega_\3$ are the volume-forms
of the unit AdS$_\3$ and three spheres:
\bea
&&\epsilon_\3= \cosh\rho\,\sinh\rho\,\, dt\wedge d\rho\wedge
d\gamma\,,\quad
\Omega_\3 =\cos\theta\,\sin\theta\, d\psi\wedge d\theta\wedge
d\phi\,,\nn\\
&&
\wtd \Omega_\3=\cos\td\theta\,\sin\td\theta\, 
d\td\psi\wedge d\td\theta\wedge
d\td\phi\,.\label{volumeforms}
\eea
After rotating coordinates as follows,
\be
\psi=\cos \alpha\ \psi_1+\sin \alpha\ \psi_2,\ \ \ \ \ {\wtd
\psi}=-\sin \alpha\ \psi_1+\cos \alpha\ \psi_2,\label{rot}
\ee
the Penrose limit can be taken to be
\bea
&&\rho\longrightarrow \fft{\rho}{\lambda}\,,\qquad
\theta\longrightarrow \fft{\theta}{\lambda}\,,\qquad 
{\wtd \theta}\longrightarrow \fft{{\wtd \theta}}{\lambda}\,,\nn\\
&&t= x^+ +\fft{x^-}{\lambda^2}\,,
\qquad \psi_1= \frac{1}{\sqrt{2}}(x^+ -\fft{x^-}{\lambda^2})\,,
\qquad \psi_2 \longrightarrow \fft{\psi_2}{\lambda}\,,
\label{limit}
\eea
with the constant $\lambda$ sent to infinity.  By this means, the
solution becomes the pp-wave
\bea
ds_{10}^2&=&- 2 dx^+\, dx^-  + H\, d{x^+}^2 + d z_i^2\,,\nn\\
F_\3 &=& dx^+ \wedge \Phi_\2\,,\label{pp}
\eea
where $\Phi_\2$ is a constant 2-form, given by
\be
\Phi_\2 =(2dz_1\wedge dz_2+\sqrt2\,\cos \alpha\ dz_3\wedge
dz_4-\sqrt2\,\sin \alpha\ dz_5\wedge dz_6)\,,\label{2form}
\ee
and 
\be
H =-\sum_{i=1}^8\mu_i^2\, z_i^2 \equiv
-(z_1^2+z_2^2)-\ft12 \cos^2 \alpha\, (z_3^2+z_4^2)-\ft12
\sin^2 \alpha\, (z_5^2+z_6^2)\,.\label{mudef}
\ee
The coordinates $z_i$ are defined to be
\bea
&&z_1=\ft1{\sqrt2}\,\rho \cos \gamma\,,\quad
  z_2=\ft1{\sqrt2}\,\rho \sin \gamma\,,\quad
  z_3=\theta \cos \phi\,,\quad
  z_4=\theta \sin \phi\,,\nn\\
&&z_5=\td\theta \cos \td\phi\,,\quad
  z_6=\td\theta \sin \td\phi\,,\quad
  z_7=\psi_2\,,\quad z_8=\varphi\,.
\eea
Note that it is straightforward to introduce a parameter $\mu$ by
scaling $x^+\rightarrow \mu\, x^+$ and $x^-\rightarrow x^-/\mu$.  In
this paper we shall set $\mu=1$.

      The above 5-5-1 system involves only the metric, the dilaton and
the 3-form field strength, and is thus valid for the type IIA, type
IIB and heterotic theories.  It can be further lifted to $D=11$ as the
5-5-2 system.  The near-horizon structure AdS$_3\times S^3\times
S^3\times S^1$ given in (\ref{ads3s3s3met}) can also be supported by a
4-form field strength of the type IIA theory; it is given by
\be 
F_\4=\lambda^2\, (\epsilon_\3 + \Omega_\3 + \wtd
\Omega_\3)\wedge d\varphi\,, 
\ee 
However, the brane interpretation becomes obscured in this case.  The
Penrose limit is now supported by the 4-form $F_\4=dx^+\wedge
\Phi_\2\wedge dz_8$.  Performing T-duality on the coordinate $\varphi$
leads to a solution supported by the the R-R 3-form in type IIB
theory.

\section{Supersymmetry and string actions}

\subsection{Supersymmetry}

\underline{M-theory and type II viewpoints}
\bigskip

      A simple way of studying the supersymmetry of the pp-wave
solution obtained in the previous section is to lift the solution to
$D=11$.  The resulting pp-wave is supported by the 4-form in $D=11$,
given by
\be
F_\4=dx^+\wedge\Phi_\2\wedge dz_9\,,
\ee
where $z_9$ is the 11'th direction.\footnote{If we lift the pp-wave
supported by the 4-form in $D=10$, $z_9$ is replaced $z_8$, and hence
these two solutions are totally equivalent in $D=11$.}  This 4-form
configuration is a special case of a general class of M-theory
pp-waves considered in \cite{clppenrose2} where the supersymmetry of
these solutions were discussed.  It is straightforward to apply the
general formalism of \cite{clppenrose2} to our solution.  To do so, we
introduce
\be
W = \ft12 \Phi_{ij} \Gamma^{ij9}=2\Gamma^{12} + \sqrt2\,\cos\a\, 
\Gamma^{34} - \sqrt2\, \sin\a\, \Gamma^{56}\,.
\ee
As discussed in \cite{clppenrose2}, there are 16 standard Killing
spinors associated with a generic pp-wave solution, which
satisfy the projection
\be 
\Gamma_{\!\!-}\, \epsilon=0\,.  
\ee 
In addition, since $\mu_{8,9}=0$, supernumerary Killing spinors arise if 
$W$ has zero eigenvalues, which occurs only for
\be
\cos^2\alpha = \sin^2\alpha = \ft12\,.\label{alphavalue}
\ee
Using the formalism of \cite{clppenrose2} we find that, for the above
$\alpha$, there are four supernumerary Killing spinors, all of which are 
independent of $x^+$ and $z^i$ for $i=7,8,9$.  The 16 standard Killing
spinors are also independent of $z^i$ for $i=7,8,9$, but only 4 out of the
16 are independent of $x^+$.  For other values of $\alpha$, there are
no Supernumerary Killing spinors, and all the standard Killing spinors
depend on $x^+$.  For this reason, we mainly focus on the case
with $\alpha$ given by (\ref{alphavalue}).

       Since the Killing spinors are independent of $z_{7,8,9}$, 
we can perform dimensional reduction and T-duality on
these directions without breaking any supersymmetry, even at the level
of field theory.  Thus, the above analysis of supersymmetry carries over
to the type IIA and IIB theories.

\bigskip\bigskip
\noindent{\underline{Heterotic viewpoint}}
\bigskip

       To study the number of Killing spinors of the solution in the
heterotic theory, we can use the same Killing-spinor calculations as
above.  However, we must impose an additional ten-dimensional
chirality condition on the Killing spinor:
\be
\epsilon=\Gamma_9\,\epsilon\,,
\ee
where $z_9$ is our 11'th coordinate.  This has the effect of reducing
the number of standard Killing spinors down to 8, 2 of which are
independent of $x^+$.  The number of supernumerary Killing spinors is
now 2.  If we had chosen a different convention for the chirality
projection, then there would not be any supernumerary Killing spinors
at all.  This implies that the sign choice of $F_\3$ is important for
the supernumerary Killing spinor in the heterotic theory.

\subsection{String action}

     PP-waves for which the function $H$ is quadratic in the
transverse space coordinates are of particular interest because they
provide backgrounds for string theory that are exactly solvable.  The
light-cone string action of the pp-wave arising from the Penrose limit
of $AdS_5\times S^5$ was studied in \cite{met,bermalnas}.  The action
associated with the Penrose limit of the D1/D5 system can be found in
\cite{bermalnas,bur1}.  In \cite{clps}, the Green-Schwarz action for
type IIA and IIB strings in an arbitrary bosonic background was
derived, in component form up to second-order in the fermionic
coordinates.  From these, type IIA and IIB string actions were
obtained in the light-cone gauge on the background of a large class of
pp-wave solutions \cite{clppenrose1,clppenrose2}.  Applying the
formalism in \cite{clppenrose1,clppenrose2} to our example, one finds
that the solution supported purely by R-R fields is particularly
simple.  The bosonic string action is given by
\be
{\cal L}_{B} =-\ft12 \dot z^2 -\ft12 z'^2 -\ft12\mu_i^2\, z_i^2\,,
\ee
where the masses of the bosonic fields $\mu_i$'s are defined in
(\ref{mudef}), given by
\be
\mu_i^2=\{1, 1, 
\ft14\, \ft14\, \ft14\,\ft14,0,0\}\,.
\label{muexp}
\ee
For the solution
supported by the type IIA 4-form field strength, the fermionic
Lagrangian is given by
\be
{\cal L}_{F} = \bar \Psi \Gamma_{\!\! +} (\not \!\del + \mu
\varrho_0\, \Gamma_8\, W)\Psi\,,
\ee
where $\varrho_i$ with $i=0,1,2$ are the world-sheet Dirac matrices,
acting on the upper and lower 16 components of the column vector
$\Psi$, and $\varrho_2$ is the chirality operator.  The fermionic
Lagrangian associated with the type IIB solution supported by the
R-R 3-form has a similar structure, given by 
\be
{\cal L}_{F} = \bar
\Psi \Gamma_{\!\! +} (\not \!\del + \mu \varrho_2\, \,
W)\Psi\,.
\ee

     As discussed in \cite{bermalnas,clppenrose1,clppenrose2}, the
existence of $x^+$-independent supernumerary Killing spinors ensures
that the corresponding string action has linearly-realised
supersymmetry.  In our case, the masses of the bosonic fields $z_i$
are given by (\ref{muexp}).  This precisely matches the masses of the
fermonic fields, given by the eigenvalues of $W$.  This is consistent
with the supersymmetry.  If we choose a different value of the
rotation angle $\alpha$, two bosonic scalars remain massless, whilst
all the fermions become massive, thereby breaking the supersymmetry.
Indeed, for generic values of $\alpha$, there are no supernumerary
Killing spinors.

       For solutions supported by the NS-NS 3-form, the string action
is slightly more complicated, since the NS-NS 3-form couples to the
worldsheet as well.  It is given by
\bea
{\cal L} = \sum_{i=1}^8 (\ft12\dot z - \ft12 z'^2 +
\ft12\mu \Phi_{ji}\,z^j z_i' -\ft12 \mu_i^2\, z_i^2) +
\bar \Psi\, (\not\!\del + \ft14 \varrho\, W)\Psi\,.
\eea

\section{T-duality and $S^1$-wrapped D3-brane}

       The pp-wave solution in $D=10$ has an eight-dimensional
transverse space.  In our case, the function $H$ is independent of the
coordinates $z_{7,8}$.  Thus, the ``natural'' transverse space of our
solution is six-dimensional.  We can perform T-duality and S-duality
to relate our solution to a D3-brane whose transverse space has six
dimensions.

      First, we perform T-duality along the $\partial/\partial x^+$
Killing direction and obtain an NS-NS string solution given by
\bea
ds_{10}^2&=&H^{-3/4}(-dt^2 +(dx+\cA_\1)^2)+H^{1/4}dz_i^2,\nn\\
F_\3^{\rm NS} &=& d(H^{-1}\,dt\wedge (dx+\cA_\1)) +
\mu\, (dx+\cA_\1)\wedge \Phi_\2\,,\nn\\
d\cA_\1 &=& \Phi_\2\,,\qquad \phi=\ft12 {\rm log}H\,.
\eea
Note that, in this solution, the worldsheet coordinate $x$ is fibred
over the transverse space.  This type of $S^1$-fibred string solution
was first obtained in \cite{overlap}.  Next, we perform S-duality such
that the string is supported by the R-R 3-form of type IIB theory.
T-dualizing the $z_{7,8}$ directions yields an $S^1$-wrapped D3-brane
given by
\bea
ds_{10}^2 &=&
H^{-1/2}(-dt^2+(dx_1+\cA_\1)^2+dx_2^2+dx_3^2)+H^{1/2}\,dz_i^2,\nn\\
F_\5 &=& d(H^{-1}\,dt\wedge (dx_1+\cA_\1)\wedge dx_2 \wedge dx_3)+
\nn\\ 
&&\mu (dx_1+\cA_\1)\wedge dx_2 \wedge dx_3 \wedge \Phi_\2+
{\rm dual},\\ 
d\cA_\1 &=& \Phi_\2\,.\nn
\eea
An $S^1$-wrapped D3-brane was first constructed in \cite{wrap} in
order to resolve the singularity while maintaining the supersymmetry
of the associated D3-brane on the resolved conifold.  Here, when the
D3-brane charge is turned on, the more general solution for $H$ is
given by
\be
H=1 + \fft{Q}{r^4} -\sum_{i=1}^6 \mu_i^2\, z_i^2\,,
\ee
where $r^2=z^iz^i$.  The terms associated with $\mu_i$ break the
conformal symmetry.  They introduce a naked singularity at some finite
$r$, which can be argued \cite{clppenrose1} to be associated with a
phase transition from type IIB theory to type IIB$^*$ theory,
introduced in \cite{hull}.  The solution has 4 Killing spinors for
non-vanishing $Q$.  For $Q=0$, or in the limit $r\rightarrow \infty$,
4 additional Killing spinors emerge.

\section{Further examples}

     In this section, we consider further examples of non-standard
intersections that give rise to AdS structure and study their Penrose
limits.  These examples can be generated from the above AdS$_3\times
S^3\times S^3$ by noting that AdS$_3$ and $S^3$ can be (locally)
expressed as a $U(1)$ bundle over AdS$_2$ and $S^2$ respectively, {\it
i.e.}
\bea
ds_{{\sst{\rm AdS}_3}}^2 &=& \ft14 (dz + \cA_\1)^2 + \ft14
ds_{{\sst{\rm AdS}_2}}^2\,,\qquad
d\cA_\1=\epsilon_\2\,,\nn\\
d\Omega_3^2 &=& \ft14 (dz + {\cal B}_\1)^2 + \ft14
d\Omega_2^2\,,\qquad d{\cal B}_\1=\Omega_\2\,,
\eea
where $\epsilon_\2$ and $\Omega_\2$ are the volume-forms of the unit
AdS$_2$ and $S^2$ respectively.  The bundle can be naturally generated
by considering a wave in AdS$_3$ or a Taub-NUT in $R^4$.  From the
$D=10$ perspective, the wave and NUT in eleven dimensions become a
D0-brane and D6-brane respectively.  The intersecting systems of this
section can be found in \cite{boonstra}.

\subsection{Type IIA examples}

\underline{D0/D4/D4/NS1}
\bigskip

The solution for the D0/D4/D4/NS1 system is given by \cite{boonstra}
\bea
ds_{10}^2 &=& H_0^{-7/8}H_4^{-3/8}{\wtd H}_4^{-3/8}H_1^{-3/4}(-dt^2+
H_0H_4{\wtd H}_4 dx^2+\nn\\ & &
H_0H_4 H_1 dy_i^2+H_0{\wtd H}_4H_1 d{\wtd y}_i^2)\,,\nn\\
F_\4 &=& {\rm e}^{-\ft12 \phi} {\ast({\wtd H_4}\ dt\wedge d^4 \td y
\wedge dH_4^{-1})}+ {\rm e}^{-\ft12 \phi}{\ast(H_4\ dt\wedge d^4 y
\wedge d{\wtd H_4}^{-1})}\,,\nn\\
F_\3 &=& dt\wedge dx\wedge dH_1^{-1}\,,\qquad F_\2=dt\wedge
dH_0^{-1}\,,\\
\phi&=& -\frac{1}{4}{\rm log}[H_4{\wtd H}_4 H_1^2/H_0^3]\,,\qquad
H_0=H_1 =H_4\,{\wtd
H}_4\,,\nn
\eea
where $H_4=H_4(y_i)$ and ${\wtd H}_4={\wtd H}_4({\wtd y}_i)$. 
The solution can be represented diagrammatically as follows:

\bigskip\bigskip
\centerline{
\begin{tabular}{c|ccccccccccc}
&$t$ & $x$ & $y_1$ & $y_2$ & $y_3$ & $y_4$ & ${\wtd y}_1$ & 
${\wtd y}_2$ & ${\wtd y}_3$ & ${\wtd y}_4$ & \\ \hline
D0&$\times$ & $-$ & $-$ & $-$ & $-$ & 
$-$ & $-$ & $-$ & $-$ & $-$ & $H_0$ \\
D4&$\times$ & $-$ & $\times$ & $\times$ & $\times$ & $\times$ & 
$-$ & $-$ & $-$ & $-$ & ${\wtd H}_4$ \\
D4&$\times$ & $-$ & $-$ & $-$ & $-$ & $-$ & 
$\times$ & $\times$ & $\times$ & $\times$ & $H_4$ \\
NS1&$\times$ & $\times$ & $-$ & $-$ & $-$ & $-$ & 
$-$ & $-$ & $-$ & $-$ & $H_1$ \\
\end{tabular}}
\bigskip

\centerline{Diagram 4. D0/D4/D4/NS1 system}
\bigskip\bigskip

We parametrize the coordinates as $dy_i^2=dy^2+y^2d\Omega_3^2$ and
$d{\wtd y}_i^2=d{\wtd y}^2+{\wtd y}^2d{\wtd \Omega}_3^2$, so that
$H_4$ and ${\wtd H}_4$ are given by (\ref{H4}). For simplicity, we
have equal charges. The near-horizon geometry is AdS$_2 \times S^3
\times S^3 \times T^2$ \cite{boonstra}, namely
\bea
ds_{10}^2 &=& \lambda^2\,( \ft18ds_{\sst{\rm AdS_2}}^2 + d\Omega_3^2 + 
d\wtd\Omega_3^2) + d\varphi_1^2 + d\varphi_2^2\,,\nn\\
F_\4 &=& 2\lambda^2\, (\Omega_\3 + \wtd \Omega_3)\wedge d\varphi_1\,,
\quad F_\3 = \ft{\lambda}{\sqrt8}\, \epsilon_\2\wedge d\varphi_1\,,\quad
F_\2= \ft{\lambda}{\sqrt8}\, \epsilon_\2\,,
\eea
Performing the Penrose limit, we have
\bea
ds_{10}^2 &=& - 2 dx^+\, dx^-  + H\, d{x^+}^2 + d z_i^2\,,\nn\\ 
F_\4 &=& dx^+ \wedge \Phi_\3\,,\quad
F_\3 = dx^+ \wedge \Phi_\2\,,\quad
F_\2 = dx^+ \wedge \Phi_\1\,,\label{pp2}
\eea
where the constant form fields are given by
\bea
\Phi_\3 &=& \ft1{\sqrt2}
(\cos \alpha\, dz_2\wedge dz_3-\sin \alpha\, dz_4\wedge
dz_5)\wedge dz_7\,,\nn\\
\Phi_\2 &=& dz_7 \wedge dz_1\,,\quad \Phi_\1 = dz_1\,,
\label{forms}
\eea
and 
\be
H =-z_1^2-\ft18 \cos^2 \alpha\,
(z_2^2+z_3^2)-\ft18
\sin^2 \alpha\, (z_4^2+z_5^2)\,.
\ee
It is worth observing that in this pp-wave solution, all the form
fields of the type IIA theory are turned on.

    Note that we can T-dualize along the $\partial/\partial x^+$
Killing direction and obtain a deformed $S^1$-wrapped type IIB NS
string solution.  Then, by T-dualizing along $z_8$ and lifting to
$D=11$, one obtains a deformed $T^2$-wrapped M2-brane solution of
M-theory given by
\bea
ds_{11}^2 &=& H^{-2/3}(-dt^2+(dx_1+\cA_\1)^2+
(dx_2+\cB_\1)^2)+H^{1/3}dz_i^2,\nn\\
F_\4 &=& d\Big(H^{-1}\, dt\wedge (dx_1+\cA_\1)\wedge (dx_2 + {\cal
B}_\1)\Big)+[\Phi_\3 \wedge dz_8+\hbox{8-dim dual}]\,,\nn\\
d\cA_\1 &=& \Phi_\2\,,\qquad d\cB_\1=\Phi_\1 \wedge dz_8\,,
\eea

\bigskip\bigskip
\noindent{\underline{D2/D4/D6/NS5}}
\bigskip

The D2/D4/D6/NS5 solution is given by \cite{boonstra}
\bea
ds_{10}^2 &=& H_2^{-5/8}H_4^{-3/8}H_6^{-1/8}H_5^{-1/4}(-dt^2+dx_1^2
+H_4H_5dx_2^2+H_2H_6H_5dy_i^2+H_2H_4dz_j^2)\,,\nn\\ 
F_\4 &=& dt\wedge dx_1\wedge dx_2\wedge dH_2^{-1}+
{\rm e}^{-\ft12 \phi} {\ast( \ dt\wedge dx_1 \wedge d^3 y
\wedge dH_4^{-1})}\,,\\
F_\3 &=& {\rm e}^{\phi}\ast (dt\wedge dx_1\wedge d^4 z\wedge
dH_5^{-1})\,,\quad   
F_\2 = {\rm e}^{-\ft32 \phi}\ast (dt\wedge dx_1\wedge dx_2\wedge d^4z\wedge
dH_6^{-1})\,,\nn\\
\phi &=& \frac{1}{4}{\rm log}[H_2 H_5^2/(H_4 H_6^3)]\,,\qquad
H_2=H_4(z_j)H(y_i),\nn
\eea
where $H_4=H_4(z_j)$, $H_6=H_5 =H(y_i)$. The solution can be
represented diagrammatically as follows:

\bigskip\bigskip
\centerline{
\begin{tabular}{c|ccccccccccc}
&$t$ & $x_1$ & $x_2$ & $y_1$ & $y_2$ & $y_3$ & $z_1$ & 
$z_2$ & $z_3$ & $z_4$ & \\ \hline
D2&$\times$ & $\times$ & $\times$ & $-$ & $-$ & $-$ & 
$-$ & $-$ & $-$ & $-$ & $H_2$ \\
D4&$\times$ & $\times$ & $-$ & $\times$ & $\times$ & $\times$ & $-$ & 
$-$ & $-$ & $-$ & $H_4$ \\
D6&$\times$ & $\times$ & $\times$ & $-$ & $-$ & $-$ & $\times$ & 
$\times$ & $\times$ & $\times$ & $H_6$ \\
NS5&$\times$ & $\times$ & $-$ & $-$ & $-$ & $-$ & $\times$ & 
$\times$ & $\times$ & $\times$ & $H_5$ \\
\end{tabular}}
\bigskip

\centerline{Diagram 2. The D2/D4/D6/NS5 system}
\bigskip\bigskip

     The near-horizon geometry is AdS$_3 \times S^3 \times S^2 \times
T^2$ \cite{boonstra}, given by
\bea
ds_{10}^2 &=& \lambda^2\, (\ft12 dS_{\sst{\rm AdS_3}}^2 +
d\Omega_3^2 + \ft14 d\wtd\Omega_2^2) + d\varphi_1^2 + d\varphi_2^2\,,\nn\\
F_\4 &=& \lambda^2 (\epsilon_\3 + 2\Omega_\3)\wedge d\varphi_1\,,
\qquad
F_\3 = \ft12\lambda\, \wtd\Omega_\2\wedge d\varphi_1\,,\qquad
F_\2 =\ft12\lambda\, \wtd\Omega_\2\,.
\eea
The Penrose limit is given by
\bea
ds_{10}^2&=&- 2\, dx^+\, dx^-  + H\, d{x^+}^2 + d z_i^2\,,\nn\\
F_\4 &=& dx^+\wedge \Phi_\3 \,,\quad
F_\3 = dx^+\wedge \Phi_\2 \,,\quad
F_\2 = dx^+ \wedge \Phi_\1 \,,
\eea
where the constant form fields are given by
\bea
\Phi_\3 &=& (2\,dz_1\wedge dz_2 + \sqrt2 \cos\alpha dz_3\wedge
dz_4) \wedge dz_6\,,\nn\\
\Phi_\2 &=& \sqrt2\, \sin\a\, dz_5\wedge dz_6\,,\quad
\Phi_\1=\sqrt2\, \sin\a\, dz_5\,,
\eea
and 
\be
H =-(z_1^2+z_2^2)-\ft12 \cos^2\alpha\,
(z_3^2+z_4^2)-2 \sin^2 \alpha\, z_5^2\,.
\ee
As in the previous example, by T-duality this solution is related to a
deformed $T^2$-wrapped M2-brane solution of M-theory.

\bigskip\bigskip
\noindent{\underline{D2/D4/D4/NS1/NS5/NS5}}
\bigskip

     The maximal number of intersecting components that give rise to an AdS
structure is six, namely the D2/D4/D4/NS1/NS5/NS5 solution, which is given
by
\cite{boonstra}
\bea
ds_{10}^2 &=& H_2^{-5/8}H_4^{-3/8}{\wtd
H}_4^{-3/8}H_1^{-3/4}H_5^{-1/4}
{\wtd H}_5^{-1/4}(-dt^2+H_2H_4{\wtd H}_4 dx_1^2+H_4H_1 {\wtd
H}_5dx_2^2+\nn\\ & &
{\wtd H}_4H_1 H_5dx_3^2+H_2H_4H_1 H_5dy_i^2+
H_2{\wtd H}_4H_1 {\wtd H}_5d{\wtd y}_i^2)\,,\nn\\
F_\4 &=& dt\wedge dx_2\wedge dx_3\wedge dH_2^{-1}+
{\rm e}^{-\ft12 \phi} {\ast({\wtd H_4}\ dt\wedge dx_3 \wedge d^3 \td y
\wedge dH_4^{-1})}+\nn\\ & &
{\rm e}^{-\ft12 \phi}{\ast(H_4\ dt\wedge dx_2\wedge d^3 y \wedge d{\wtd
H_4}^{-1})}\,,\\
F_\3 &=& dt\wedge dx_1\wedge dH_1^{-1}+
{\rm e}^{\phi}\ast ({\wtd H}_5 dt\wedge dx_1\wedge dx_2\wedge d^3 {\wtd 
y}\wedge dH_5^{-1})+\nn\\ & &
{\rm e}^{\phi}\ast (H_5 dt\wedge dx_1\wedge dx_3\wedge d^3 y\wedge
d{\wtd H}_5^{-1})\,,\nn\\
\phi &=& \frac{1}{4}{\rm log}[H_2 (H_5{\wtd
H}_5)^2/(H_1^2 H_4 {\wtd H}_4)]\,,\qquad
H_2=H_1=H(y_i){\wtd H}({\wtd y}_i),\nn
\eea
where $H_4=H_5=H(y_i)$ and ${\wtd H}_4={\wtd H}_5={\wtd H}({\wtd
y}_i)$. The solution can be represented diagrammatically as follows:

\bigskip\bigskip
\centerline{
\begin{tabular}{c|ccccccccccc}
&$t$ & $x_1$ & $x_2$ & $x_3$ & $y_1$ & $y_2$ & $y_3$ & 
${\wtd y}_1$ & ${\wtd y}_2$ & ${\wtd y}_3$ & \\ \hline
D2&$\times$ & $-$ & $\times$ & $\times$ & $-$ & 
$-$ & $-$ & $-$ & $-$ & $-$ & $H_2$ \\
D4&$\times$ & $-$ & $-$ & $\times$ & $-$ & $-$ & 
$-$ & $\times$ & $\times$ & $\times$ & $H_4$ \\
D4&$\times$ & $-$ & $\times$ & $-$ & $\times$ & $\times$ & 
$\times$ & $-$ & $-$ & $-$ & ${\wtd H}_4$ \\
NS1&$\times$ & $\times$ & $-$ & $-$ & $-$ & $-$ & 
$-$ & $-$ & $-$ & $-$ & $H_1$ \\
NS5&$\times$ & $\times$ & $\times$ & $-$ & $-$ & $-$ & 
$-$ & $\times$ & $\times$ & $\times$ & $H_5$ \\
NS5&$\times$ & $\times$ & $-$ & $\times$ & $\times$ & $\times$ & 
$\times$ & $-$ & $-$ & $-$ & ${\wtd H}_5$ \\
\end{tabular}}
\bigskip

\centerline{Diagram 5. D2/D4/D4/NS1/NS5/NS5 system}
\bigskip\bigskip

We parametrize the coordinates as $dy_i^2=dy^2+y^2d\Omega_2^2$ and
$d{\wtd y}_i^2=d{\wtd y}^2+{\wtd y}^2d{\wtd \Omega}_2^2$, so that 
\be
H_4=H_5=1+\frac{\lambda}{y},
\ee
and likewise for ${\wtd H}_4$ and ${\wtd H}_5$. The near-horizon geometry
is $AdS_2 \times S^2 \times S^2 \times T^4$ \cite{boonstra}. 
Explicitly, we have
\bea
ds_{10}^2 &=& \ft14\lambda^2\,\Big( \ft12 ds_{\sst{\rm AdS_2}}^2 +
d\Omega_\2^2 + d\wtd \Omega_2^2\Big) + d\varphi_1^2
+d\varphi_2^2 + d\varphi_3^2 + d\varphi_4^2\,,\nn\\ 
F_\4 &=& \ft12\lambda\, (\Omega_\2\wedge d\varphi_1\wedge d\varphi_3 +
\wtd \Omega_\2\wedge d\varphi_1\wedge d\varphi_4 +
\ft1{\sqrt2}\epsilon_\2\,\wedge d\varphi_3\wedge d\varphi_4)\,,\nn\\
F_\3 &=& \ft12\lambda\, (\Omega_\2\wedge d\varphi_4 + \wtd\Omega_2
\wedge d\varphi_3 + \ft1{\sqrt2}\, \epsilon_\2\wedge d\varphi_4)\,.
\eea
We find that the Penrose limit of this system is given by
\bea
ds_{10}^2 &=& - 2 dx^+\, dx^-  + H\, d{x^+}^2 + d z_i^2\,,\nn\\
F_\4 &=& dx^+ \wedge \Phi_\3\,,\quad F_\3=dx^+ \wedge \Phi_\2\,,\label{pp3}
\eea
where the constant form fields are given by
\bea
\Phi_\3 &=& \sqrt{2}\, (dz_6\wedge dz_7\wedge dz_1+
\ft12 \cos \alpha\, dz_2\wedge
dz_5\wedge dz_6-\ft12 \sin \alpha\, dz_3
\wedge dz_5\wedge dz_7)\,,\nn\\
\Phi_\2 &=& \sqrt{2}\, (dz_5\wedge dz_1+
\ft12\cos \alpha\, dz_2\wedge dz_7
-\ft12\sin \alpha\, dz_3\wedge dz_6\,,\label{forms2}
\eea
and
\be
H = -z_1^2-\ft12 \cos^2 \alpha\,
z_2^2-\ft12 \sin^2 \alpha\, z_3^2\,.
\ee
Note that this solution can be easily lifted to $D=11$ as the Penrose
limit of the\\ M2/M2/M5/M5/M5/M5 system.  There are six terms in the
4-form field stength in the 11-dimensional pp-wave solution.  This is
a special case of the general class of M-theory pp-wave solution
obtained in \cite{clppenrose2}.

\subsection{Type IIB examples}

\underline{D3/D1/D5/NS1/NS5}
\bigskip

    The near-horizon geometry of this system is AdS$_2\times S^3\times
S^2\times T^3$ \cite{boonstra}.  Explicitly we find that
\bea
ds_{10}^2&=&\lambda^2 (\ft18\, ds_{\sst{{\rm AdS}_2}}^2 +
d\Omega_3^2 + \ft14 d\wtd\Omega_2^2) + d\varphi_1^2 +
d\varphi_2^2 + d\varphi_3^2\,,\nn\\
F_\5&=&2\lambda^2\, \Omega_\3\wedge d\varphi_1\wedge d\varphi_3 + {\rm
dual}\,,\nn\\
F_\3^{\rm RR} &=& \ft12\lambda\, (\wtd \Omega_\2\wedge d\varphi_1 +
\ft1{\sqrt2}\, \epsilon_\2\wedge d\varphi_3)\,,\\
F_\3^{\rm NS} &=& \ft12\lambda\, (\wtd \Omega_\2\wedge d\varphi_3 +
\ft1{\sqrt2}\, \epsilon_\2\wedge d\varphi_1)\,.\nn
\eea
The Penrose limit is given by
\bea
ds_{10}^2 &=& -2\, dx^+ dx^-+H d{x^+}^2+dz_i^2\,,\nn\\
F_\5 &=& dx^+ \wedge \Phi_\4\,,\quad
F_\3^{RR}=dx^+ \wedge \Phi_\2^{RR}\,,\quad
F_\3^{NS}=dx^+ \wedge \Phi_\2^{NS}\,,
\eea
where the constant form fields are given by
\bea
\Phi_\4 &=& \ft{1}{\sqrt2}\,\cos \alpha\, dz_2\wedge dz_3\wedge
dz_6\wedge dz_8\,,\nn\\
\Phi_\2^{RR} &=& dz_1\wedge dz_8-\ft1{\sqrt2} \sin \alpha\,
dz_4\wedge dz_6\,,\\
\Phi_\2^{NS} &=& dz_1\wedge dz_6-\ft1{\sqrt2} \sin \alpha\,
dz_4\wedge dz_8)\,,\nn
\eea
and 
\be
H=-z_1^2-\ft18 \cos^2 \alpha\, (z_2^2+z_3^2)-\ft12 \sin^2 \alpha\, z_4^2.
\ee

\bigskip\bigskip
\noindent{\underline{D3/D5/D5/NS5/NS5}}
\bigskip

         The near-horizon geometry of this system is AdS$_3\times
S^2\times S^2\times T^3$ \cite{boonstra}.  Explicitly we have
\bea
ds_{10}^2 &=& \lambda^2\,  (\ft12 ds_{\sst{{\rm AdS}_3}}^2 + \ft14
d\Omega_2^2 + \ft14 d\wtd \Omega_2^2) + d\varphi_1^2 +
d\varphi_2^2 + d\varphi_3^2\,,\nn\\
F_\5 &=& \lambda^2 \epsilon_\3\wedge d\varphi_1\wedge \varphi_3
+ {\rm dual}\,,\nn\\
F_\3^{\rm RR} &=& \ft12\lambda\, (\Omega_\2\wedge d\varphi_1 +
\wtd \Omega_\2\wedge d\varphi_3)\,,\\
F_\3^{\rm NS} &=& \ft12\lambda\, (\Omega_\2\wedge d\varphi_3 +
\wtd \Omega_\2\wedge d\varphi_1)\,,\nn
\eea
The Penrose limit is given by
\bea
ds_{10}^2 &=& -2\, dx^+ dx^-+H d{x^+}^2+dz_i^2\,,\nn\\
F_\5 &=& dx^+ \wedge \Phi_\4\,,\quad
F_\3^{RR}=dx^+ \wedge \Phi_\2^{RR}\,,\quad
F_\3^{NS}=dx^+ \wedge \Phi_\2^{NS}\,,
\eea
where the constant form fields are given by
\bea
\Phi_\4 &=& 2 dz_1\wedge dz_2\wedge dz_6\wedge dz_8\,,\nn\\
\Phi_\2^{RR} &=& \sqrt2\,(\cos \alpha\, dz_3\wedge dz_6-
\sin \alpha\, dz_4\wedge dz_8)\,,\\
\Phi_\2^{NS} &=& \sqrt2\,(\cos \alpha dz_3\wedge dz_8-\sin
\alpha\, dz_4\wedge dz_6)\,,\nn
\eea
and 
\be
H=-(z_1^2+z_2^2)-2 \cos^2 \alpha\, z_3^2-2 \sin^2 \alpha\,
z_4^2.
\ee

\section{Conclusions}

     We have considered the Penrose limit of $AdS_3$ and $AdS_2$ spacetimes
that arise from various non-standard brane intersections in type IIA, IIB
and heterotic theories.  We have focused on the non-standard intersection
of two 5-branes and one string, whose near-horizon geometry is
AdS$_3\times S^3\times S^3\times S^1$.  A new feature is that the product 
spacetime involves two 3-spheres.  After making an orthogonal rotation of
the two foliating circles of the two 3-spheres, we obtain a one-parameter
family of Penrose limits.  Through the study of the supersymmetry of the
resulting pp-wave, we find that supernumerary Killing spinors arise only
for a 45 degree rotation.  Thus, there are 4 supernumerary Killing spinors
in addition to the 16 standard Killing spinors.  The existence of the
supernumerary Killing spinors ensures that the corresponding light-cone
string action has linearly-realised supersymmetry.  It is of interest to
further investigate the role of the variable rotation parameter for 
supersymmetry breaking.

\section*{Acknowledgements}

We would like to thank Mirjam Cveti\v{c} and Chris Pope for useful
conversations.


\begin{thebibliography}{99}

\bibitem{kg} J. Kowalski-Glikman,
{\it Vacuum States In Supersymmetric Kaluza-Klein Theory,}
Phys. Lett. {\bf B134}, 194 (1984).

\bm{blafighulpap} M. Blau, J. Figueroa-O'Farrill, C. Hull and
G. Papadopoulos, {\it A new maximally supersymmetric background of IIB
superstring theory}, JHEP {\bf 0201} (2002) 047, hep-th/0110242.

\bibitem{blafighulpap2} M. Blau, J. Figueroa-O'Farrill, C. Hull and
G. Papadopoulos, {\it Penrose limits and maximal supersymmetry},
hep-th/0201081.

\bm{met} R.R. Metsaev,{\it Type IIB Green-Schwarz superstring in plane
wave Ramond-Ramond background}, Nucl. Phys. {\bf B625}, 70 (2002),
hep-th/0112044.

\bm{bermalnas} D. Berenstein, J. Maldacena and H. Nastase, {\it
Strings in flat space and pp waves from N = 4 super Yang Mills},
hep-th/0202021.

\bm{bur1}
R.R. Metsaev and A.A. Tseytlin,
{\it Exactly solvable model of superstring in plane wave
Ramond-Ramond  background}, hep-th/0202109.

\bm{bur1.5}
M. Blau, J. Figueroa-O'Farrill and G. Papadopoulos,
{\it Penrose limits, supergravity and brane dynamics},
hep-th/0202111.

\bm{bur2}
N. Itzhaki, I.R. Klebanov and S. Mukhi,
{\it PP wave limit and enhanced supersymmetry in gauge theories},
hep-th/0202153.

\bm{bur3}
J. Gomis and H. Ooguri,
{\it Penrose limit of $N = 1$ gauge theories}, hep-th/0202157.

\bm{bur4}
J.G. Russo and A.A. Tseytlin,
{\it On solvable models of type IIB superstring in NS-NS and
R-R plane wave  backgrounds}, hep-th/0202179.

\bm{bur5}
L.A. Pando-Zayas and J. Sonnenschein,
{\it On Penrose limits and gauge theories}, hep-th/0202186.

\bm{bur5.5}
M. Alishahiha and M.M. Sheikh-Jabbari,
{\it The pp-wave limits of orbifolded AdS$_5\times S^5$},
hep-th/0203018.

\bm{bur6}
M. Billo' and I. Pesando,
{\it Boundary states for GS superstrings in an Hpp wave background},
hep-th/0203028.

\bm{bur7}
N. Kim, A. Pankiewicz, S-J. Rey and S. Theisen,
{\it Superstring on pp-wave orbifold from large-$N$ quiver gauge theory},
hep-th/0203080.

\bibitem{clppenrose1} M. Cveti\v c, H. L\"u and C.N. Pope,
{\it Penrose limits, pp-waves and deformed M2-branes,} hep-th/0203082.

\bibitem{bur7.5} T. Takayanagi and S. Terashima,
{\it Strings on orbifolded pp-waves,} hep-th/0203093.

\bibitem{bur8} U. Gursoy, C. Nunez and M. Schvellinger,
{\it RG flows from Spin(7), CY 4-fold and HK manifolds to AdS,
Penrose  limits and pp waves,} hep-th/0203124.

\bibitem{bur9} E. Floratos and A. Kehagias,
{\it Penrose limits of orbifolds and orientifolds,} hep-th/0203134.

\bibitem{mich}
J. Michelson,
{\it (Twisted) toroidal compactification of pp-waves,}
hep-th/0203140.

\bibitem{Gueven}
R. Gueven,
{\it Randall-Sundrum zero mode as a Penrose limit,}
hep-th/0203153.

\bibitem{Das:2002cw}
S.R. Das, C. Gomez and S.J. Rey,
{\it Penrose limit, spontaneous symmetry breaking and holography in
pp-wave  background,} hep-th/0203164.

\bibitem{bur10} C.S. Chu and P.M. Ho,
{\it Noncommutative D-brane and open string in pp-wave background with
$B$-field}, hep-th/0203186.

\bibitem{clppenrose2}
M. Cveti\v c, H. L\"u and C.N. Pope,
{\it M-theory PP-Waves, Penrose Limits and Supernumerary Supersymmetries,}
hep-th/0203229.

\bibitem{bur11}
D. Berenstein, E. Gava, J. Maldacena, K.S. Narain and H. Nastase,
{\it Open strings on plane waves and their Yang-Mills duals,}
hep-th/0203249.

\bibitem{bur12}
J.P. Gauntlett and C.M. Hull,
{\it PP-waves in 11-dimensions with extra supersymmetry,}
hep-th/0203255.

\bibitem{bur13}
P. Lee and J.W. Park,
{\it Open strings in PP-wave background from defect conformal field theory,}
hep-th/0203257.

\bm{khuri} R.R. Khuri, {\it A comment on the string solitons},
Phys. Rev. {\bf D48} (1993) 2947, hep-th/9305143.

\bm{bbj} K. Behrndt, E. Bergshoeff and B. Janssen, {\it Intersecting
D-branes in ten and six dimensions}, Phys. Rev. {\bf D55} 3785
(1997), hep-th/9604168.

\bm{gkt} J.P. Gauntlett, D.A. Kastor and J. Traschen, {\it Overlapping
branes in M-theory}, Nucl. Phys. {\bf B478} (1996) 544,
hep-th/9604179.

\bm{cow} P.M. Cowdall, P.K. Townsend, {\sl Gauged supergravity vacua
from intersecting branes}, Phys. Lett. {\bf B429} (1998) 281,
Erratum-ibid. B434 (1998) 458, hep-th/9801165.

\bibitem{clps} M. Cveti\v c, H. L\"u, C.N. Pope and K.S. Stelle,
{\it T-duality in the Green-Schwarz formalism, and the
massless/massive IIA  duality map,}
Nucl.\ Phys.\ {\bf B573}, 149 (2000),
hep-th/9907202.

\bibitem{overlap}
H. L\"u and J.F. V\'azquez-Poritz,
{\it Resolution of overlapping branes,} hep-th/0202075.

\bibitem{wrap}
H. L\"u and J.F. V\'azquez-Poritz,
{\it $S^1$-wrapped D3-branes on conifolds,}
hep-th/0202175.

\bibitem{hull} C.M. Hull, {\it Timelike T-duality, de Sitter
space, large N gauge theories and topological field theory,} JHEP {\bf
9807} (1998) 021, hep-th/9806146.

\bibitem{boonstra} H.J. Boonstra, B. Peeters and K. Skenderis, {\it Brane
intersections, anti-de Sitter spacetimes and dual superconformal field
theories,} Nucl. Phys. {\bf B533} (1998) 127-162, hep-th/9803231. 

\end{thebibliography}
\end{document}